\documentstyle[aps,prl,amsmath,amssymb]{revtex}
\draft
\begin{document}
 \title{Localization corrections and 
small-q phonon-mediated unconventional superconductivity
in the cuprates}
\author{
Georgios Varelogiannis}
\address{
Institute of Electronic Structure and Laser,
Foundation for Research and Technology Hellas\\
P.O. Box 1527, GR-71110
Heraklion Crete, Greece} 
\date{\today}
\twocolumn[\hsize\textwidth\columnwidth\hsize\csname@twocolumnfalse\endcsname
\maketitle
\begin{abstract}
Taking into account the first localization corrections
in the electron-impurity self-energy we study
the effect of non-magnetic impurities on
unconventional superconductivity (SC)
mediated by small-q electron-phonon scattering.
We show that when van Hove singularities are close to the Fermi
level
making the electronic system anisotropic as
in the high-$T_c$ oxides,
both the
d-wave and s-wave states
are sensitive to non-magnetic impurities and beyond a critical
impurity concentration SC disappears
in {\it both gap symmetry channels}.
Impurity doping may induce a first order transition from
d-wave to s-wave SC, but no saturation
of the impuruty effect is reported due to the intrinsically
anisotropic character of the localization corrections
in this context.
\end{abstract}

]
\narrowtext
%\twocolumn
%

It is now established that the pairing
symmetry in the high-$T_c$ cuprates
is dominantly $d$-wave
\cite{Tsuei}. The most common approach associates the
d-wave symmetry with a pairing mechanism
involving
spin fluctuations. 
An alternative approach has been proposed recently
according to which the gap symmetry could be an energetically
{\it marginal} parameter \cite{SD}.
The basis of this last approach is the
conjencture that the pairing scattering is
dominated by {\it attractive longwavelength processes
involving phonons} \cite{SD}.
Small-q pairing may lead to d-wave 
superconductivity (SC) mediated by 
phonons in the cuprates, and this approach becomes particularly
attractive in the light of recent experimental 
reports suggesting new          
evidence for a phonon mechanism in these compounds \cite{Shen}.

When forward processes dominate the pairing,
there is a decorrelation tendency for SC in the various regions
of the Fermi surface and gap
anisotropies are driven by the anisotropies
of the electronic system itself \cite{DOSdriven}.
This situation has been named the {\it Momentum Decoupling} (MD) 
regime \cite{SD,DOSdriven}.
Depending on the details of the 
conventional
Coulomb pseudopotential one can naturally obtain 
d-wave gap symmetries
and it is even possible
to obtain gap symmetry transitions
by changing the doping or the concentration 
of impurities \cite{SD,Abrikosov}.
The presence of a significant s-wave component
in the dominantly d-wave gap of YBCO is understood in the MD regime as
resulting from the orthorhombic distortion of the $CuO_2$ planes
\cite{ortho} and in this context we can also fit the anisotropic
pressure dependence of $T_c$ in YBCO compounds \cite{PRB2000}.
Effectively longwavelength
pairing in high-$T_c$ compounds 
has been considered by many authors in different
contexts eventhough in some cases it is only implicit \cite{smallQ}.
Small-q phonon pairing has also been shown to reproduce
accurately
the X-shaped d-wave gap and related anomalies in 
$\kappa-BEDT$ organic superconductors \cite{organic}.

In the MD regime, the condensation free energies of
anisotropic s- and d-waves may have similar magnitude.
Since non-magnetic impurities 
reduce more efficiently the condensation free enrgy of a d-wave gap,
beyond a critical impurity density one expects that the s-wave
gap may have a lower free energy and thus a d-wave
to s-wave transition may be induced \cite{Abrikosov}.
As a result,     
at sufficiently high impurity concentration,
SC would be expected to be insensitive to impurities as
in conventional
s-wave superconductors following Anderson's theorem. However,
impuirities destroy superconductivity in the cuprates and no regime
of saturation of the impurity effect is observed \cite{expImp}.
We show here that the saturation regime is in fact never reached
because in such anisotropic systems 
the localization corrections corresponding to the crossed impurity
diagrams are {\it intrinsically anisotropic} and therefore
Anderson's theorem do not applies. 

Many studies of
the effects of normal
impurities in unconventional SC                 
have been carried out
especially
in heavy Fermion
\cite{HF} and 
high-$T_c$ SC \cite{Studies}.
With some exceptions like in Ref. \cite{Ferhen}, the influence          
of realistic anisotropic bands 
and in particular van Hove singularities on the impurity effects has not
been investigated in much detail.
The first impurity self-energy diagram 
considered in the usual studies of SC    
is shown in figure 1a and the corresponding self energy is given by
$$
\Sigma^{(1)} (\omega_n) =
n_i V^2 \sum_{\vec{k}} 
{1\over i\omega_n - \xi_{\vec{k}}-
\Sigma (\vec{k},\omega_n)}
\eqno(1)
$$
where $n_i$ is the density of impurities and $V$ the scattering potential.
This type of self-energy diagrams cannot lead to localization and
correspond to ladder diagrams 
in the particle-hole propagator \cite{book}.
Enhancing the concentration of impurities, the localization effects in 
the transport are known to occur from the series of maximally crossed 
diagrams \cite{book,Crossed} in the particle-hole propagator. 
In the self-energy, the localization effects manifest in the   
diagrams with crossed electron-impurity interaction lines.
We adopt here a ``perturbative'' approach including only        
the first crossed diagram and this is what we call  
the first localization correction in the self-energy. 
We will also limit our discussion
in the Born approximation. We will focus on qualitative 
points that are related to the momentum 
dependence of these self energy
corrections 
which are probably robust even if next order diagrams are 
included, because all crossed self energy 
diagrams are expected momentum dependent
in highly anisotropic electronic systems.

The first localization correction 
$\Sigma^{(2)}$ that we consider here 
is given by (Figure 1b)
$$
\Sigma^{(2)}(\vec{k},\omega_n)=
n_i^2V^4\sum_{\vec{k}',\vec{k}''}
{1\over [i\omega_n - \xi_{\vec{k}'}-\Sigma(\vec{k}',\omega_n)]}
$$
$$
\times
{1\over [i\omega_n - \xi_{\vec{k}''}-\Sigma(\vec{k}'',\omega_n)]}
\times
$$
$$
\times
{1\over i\omega_n - \xi_{\vec{k}'-\vec{k}''+\vec{k}}-
\Sigma(\vec{k}'-\vec{k}''+\vec{k},\omega_n)}
\eqno(2)
$$
$\Sigma^{(2)}$ is second order in $n_iV^2$ compared to 
$\Sigma^{(1)}$ and our approach may be viewed as a
first order expansion in $n_iV^2$
of the localization effects.
The impurity scattering potential $V$ is considered momentum 
independent 
(local) in our approach. As a result, the
$\Sigma^{(1)}$ term is also momentum independent.
On the other hand,
$\Sigma^{(2)}$ is {\it momentum dependent}
despite the local character of the 
scattering potential. Corrections of the type of $\Sigma^{(2)}$ introduce
therefore an {\it intrinsically anisotropic} impurity scattering
if the electronic system is itself anisotropic.
Notice that the anisotropic character of the localization terms in
anisotropic electronic systems has already been
pointed out in Ref.
\cite{WB}.
We therefore expect a non-trivial interplay 
of our impurity self-energy effects with 
anisotropic SC that we have explored numerically.

In the superconducting state,
the Dyson equation for the impurity self-energy effects 
reads \cite{HF}         
$$
\bigl[ i\omega_n\hat{\tau}_0-\xi_{\vec{k}}\hat{\tau}_3-\Delta_{\vec{k}}\hat{\tau}_1
-\hat{\Sigma}(\vec{k},i\omega_n)\bigr]\hat{G}(\vec{k},i\omega_n)=\hat{\tau}_0
\eqno(3)
$$
where $\hat{\tau}_i$ are the Pauli matrices in the usual notations.
%$
%hat{G}={-i\tilde{\omega}_n\hat{\tau}_0+\tilde{\Delta}_{\vec{k}}\hat{\tau}_1+
%tilde{\xi}_{\vec{k}}\hat{\tau}_3\over \tilde{\omega}_n^2-
%tilde{\Delta}_{\vec{k}}^2-\tilde{\xi}_{\vec{k}}^2}
%eqno(3)
%$
Solving the SC self-consistency problem that results from equations (1-3)
is a formidable task that requires further simplifications.
We adopt here a {\it ``generalized''} Born approximation.
Within this approximation we can write: 
$$
\tilde{\omega}_n=\omega_n+in_i V^2
\sum_{\vec{k}'}
{i\omega_n\over (i\omega_n)^2 -\Delta_{\vec{k}'}^2- \xi_{\vec{k}'}^2}
$$
$$
+
in_i^2V^4\sum_{\vec{k}',\vec{k}''}
{i\omega_n\over (i\omega_n)^2 -\Delta_{\vec{k}'}^2- \xi_{\vec{k}'}^2}
\times
$$
$$
%times
{i\omega_n\over [(i\omega_n)^2 -\Delta_{\vec{k}''}^2- \xi_{\vec{k}''}^2]}
{i\omega_n\over [(i\omega_n)^2 -\Delta_{\vec{k}'-\vec{k''}+\vec{k}}^2- 
\xi_{\vec{k}'-\vec{k}''+\vec{k}}^2]}
\eqno(4)
$$
$$
\tilde{\Delta}_{\vec{k}}=\Delta_{\vec{k}}+
n_i V^2
\sum_{\vec{k}'}
{\Delta_{\vec{k}'}\over (i\omega_n)^2 -\Delta_{\vec{k}'}^2- \xi_{\vec{k}'}^2}
$$
$$
+
n_i^2V^4\sum_{\vec{k}',\vec{k}''}
{\Delta_{\vec{k}'}
\over (i\omega_n)^2 -\Delta_{\vec{k}'}^2- \xi_{\vec{k}'}^2}
\times
$$
$$
%times
{\Delta_{\vec{k}''}
\over [(i\omega_n)^2 -\Delta_{\vec{k}''}^2- \xi_{\vec{k}''}^2]}
{\Delta_{\vec{k}'-\vec{k}''+\vec{k}}
\over [(i\omega_n)^2 -\Delta_{\vec{k}'-\vec{k}''+\vec{k}}^2-
\xi_{\vec{k}'-\vec{k}''+\vec{k}}^2]}
\eqno(5)
$$
$$
\tilde{\xi}_{\vec{k}}=\xi_{\vec{k}}
+n_i V^2
\sum_{\vec{k}'}
{\xi_{\vec{k}'}\over (i\omega_n)^2 -\Delta_{\vec{k}'}^2- \xi_{\vec{k}'}^2}
$$
$$
+
n_i^2V^4\sum_{\vec{k}',\vec{k}''}
{\xi_{\vec{k}'}
\over (i\omega_n)^2 -\Delta_{\vec{k}'}^2- \xi_{\vec{k}'}^2}
\times
$$
$$
%times
{\xi_{\vec{k}''}
\over [(i\omega_n)^2 -\Delta_{\vec{k}''}^2- \xi_{\vec{k}''}^2]}
{\xi_{\vec{k}'-\vec{k}''+\vec{k}}
\over [(i\omega_n)^2 -\Delta_{\vec{k}'-\vec{k}''+\vec{k}}^2-
\xi_{\vec{k}'-\vec{k}''+\vec{k}}^2]}
\eqno(6)
$$
and the superconducting properties are obtained solving a 
gap equation as usually
$$
\Delta_{\vec{k}}=-T\sum_{\vec{k}'}\sum_{\omega_n}
{\Lambda(\vec{k},\vec{k}')\tilde{\Delta}_{\vec{k}'}\over
\tilde{\omega}_n^2+\tilde{\Delta}_{\vec{k}'}^2+\tilde{\xi}_{\vec{k}'}^2}
\eqno(7)
$$
where $\Lambda(\vec{k},\vec{k}')$ is the pairing potential
to be specified below and            
the fermion Matsubara frequencies are given by $\omega_n=(2n+1)\pi T$.

In the above formulation, strong-coupling 
effects in the pairing are neglected
for the sake of keeping the calculations computationally tractable.
Retardation could however interfere with the 
effects of disorder as discussed for isotropic cases in Ref. \cite{Belitz}.
The electronic dispersion $\xi_{\vec{k}}$ that
we consider is a tight-binding           
modelization of the $CuO_2$ planes
of high-$T_c$ cuprates that we have used previously in the study of the 
normal state $dc$ resistivity \cite{dc}:
$
\xi_{\vec{k}}-E_{vH}= t_1(\cos k_x+\cos k_y)+t_2 \cos k_x \cos k_y +
{1\over 2}
t_3 (\cos 2 k_x + \cos 2 k_y) +
{1\over 2}
t_4 (\cos 2k_x \cos k_y + \cos k_x \cos 2k_y)
+t_5 \cos 2k_x\cos 2k_y
%eqno(8)
$
with $t_1=-0.525$, $t_2=0.0337$, $t_3=0.0287$, $t_4=-0.175$ and $t_5=0.0175$
(the lattice spacing is taken equal to one).
This type of dispersion reproduces the {\it extended}
regions of flat bands centered
at the saddle points $(0,\pm\pi)$ and $(\pm\pi,0)$ 
and $E_{vH}$ is the distance in energy
of the van Hove peak in the angle integrated electronic density of states
from the Fermi level (when $E_{vH}=0$ the peak
in the DOS is exactly at 
the Fermi level \cite{dc}).

To study the effect of localization corrections in the MD regime we 
consider a pairing pottential similar to that in Refs. 
\cite{SD,DOSdriven,ortho,PRB2000} written for small-$\bf{q}$ as follows
$$
\Lambda(\vec{k}-\vec{p})=-\Lambda^0
\biggl( 1+{|\vec{k}-\vec{p}|^2\over Q_c^2}\biggr)^{-1}+\mu^*(\vec{q})
\eqno(8)
$$
The pairing pottential is isotropic and dominated by the processes that 
transfer a momentum smaller than $Q_c$ which plays the role of an effective
momentum cut-off and is taken here equal to $\pi/6$ \cite{DOSdriven}.
Although the pairing potential of eq. (8) is isotropic the gap obtained from equation
(7) is anisotropic reflecting mainly the anisotropy of the Fermi 
velocity in our system \cite{DOSdriven}. 
The Coulomb pseudopotential $\mu^*$ is taken 
momentum independent and equal to $\Lambda^0/10$ so that
without impurities the anisotropic d-wave 
solution is the one with the lowest free energy \cite{SD}.
We display in Figure 2 typical d-wave
and s-wave solutions that we have obtained using our kernel
and our model dispersion with               
the extended van-Hove singularities taken exactly on
the Fermi surface. In both symmetry channels the solutions
are anisotropic. The largest gap values are observed
in the areas around $(0,\pm \pi)$ and 
$(\pm \pi,0)$ where the extended flat regions of the model
dispersion are centered and the density of states is high.
Density of states driven anisotropies is a key
characteristic of the MD regime \cite{SD,DOSdriven}.

We study the evolution of the critical temperature
of such states when an increasing density of impurities
is introduced. 
We show in figure 3a a characteristic set 
of our results when the impurity potential is about half the bandwidth
and the van Hove peak at $30meV$ below the Fermi level.
One can see in figure 3a that
when the concentration of impurities grows,                 
the anisotropic s-wave state (circles) 
becomes rapidly more favorable than the anisotropic d-wave
state (triangles).
This transition 
has been predicted by Abrikosov \cite{Abrikosov}.
Such transitions from d-wave to s-wave
have also been reported as a function of the magnitude and momentum 
structure of the Coulomb pseudopotential using the 
same pairing kernel \cite{SD}. Applying a uniform magnetic field
parallel to the planes (Zeeman field) may also induce gap symmetry
transitions in this context \cite{organic}. 
This {\it volatility}
of the gap symmetry could not be obtained in the context
of a spin fluctuations pairing
mechanism.                        

Despite the fact that 
at high impurity density we are in an 
anisotropic s-wave state
the 
critical temperature is continuously reduced
when we further enhance the concentration of 
impurities. No saturation regime is reached in qualitative 
agreement with the experiments on high-$T-c$ cuprates. 
Like the d-wave state,
the s-wave state {\it collapses}      
beyond a critical concentration of impurities.
This unexpected behavior is due to    
the proximity of the van Hove singularity to the 
Fermi level as can be clearly seen in Fig. 3b where
we report results in the s-wave channel as a function
of the distance of the van Hove peak from the Fermi level.

The momentum dependence of the 
crossed diagram corrections as well as the density of states
driven anisotropic SC induced by the 
singular at small-q kernel we are using \cite{DOSdriven}
play a crucial role in eliminating the saturation of the 
impurity effect. In fact, we have performed calculations
in which the terms related to $\Sigma^{(2)}$ were set equal to 
zero and the resulting behavior is qualitatively
similar to that obtained
including $\Sigma^{(2)}$ but with the van-Hove singularities
far from the Fermi level. Moreover, the impurity effects in the
s-wave channel depend on the magnitude of $Q_c$, being more
drastic when $Q_c$ is reduced illustrating the importance of
density of states driven anisotropies in the MD regime 
\cite{DOSdriven,SD}. 

A quantitative 
comparison of our calculations with the experiments 
at that level
of approximation
is out of question and
the quantitative 
dependence of our results on the various 
parameters will be explored elsewhere.
The present results clearly demonstrate that
in anisotropic 
systems, 
the interplay of large concentrations of normal impurities and 
unconventional superconductivty may be highly 
non trivial and extrapolation
of ideas based on Anderson's theorem should be made cautiously.
When van-Hove singularities and
density of states driven anisotropies are 
in game as in the MD regime, a saturation regime of the 
impurity effects on SC is never reached.
As the density of impurities grows the inclusion of localization
corrections is unavoidable and these last corrections are
intrinsically anisotropic in this regime preventing the aplicability
of Anderson's theorem.

I acknowledge enlightening comments by
Boris Altshuler
on the aplicability of Anderson's theorem in
the presence of effectively momentum dependent scattering.

%newpage

%newpage

\vskip 0.3cm
{\Large \bf Figure Captions}
\vskip 0.2cm

{\bf Figure 1:}
Electron impurity self-energy diagrams. In (a) the self-energy effects 
considered in the usual Born approximation 
and in (b) the first localization corrections.
\vskip 0.2cm

{\bf Figure 2:} Typical d-wave (a) and s-wave (b) 
gap solutions over the first
Brillouin zone of our model dispersion 
obtained using a 
pairing potential having the form of Eq. (8) as described in the texte.
\vskip 0.2cm

{\bf Figure 3:} (a)
The ratio $T_c/T_c^d$ as a function of the density of 
impurities ($T_c^d$ is the critical temperature of the d-wave gap without 
impurities). The d-wave solution dominates without impurities.
As the density of impurities grows,
the $T_c$ of the s-wave (circles) 
becomes higher than that of the d-wave (triangles)
and s-wave becomes the physical state of the system.
In both symmetries $T_c$ is reduced continuously with impurity doping
towards zero.  
(b) Evolution of $T_c$ in the s-wave channel
for three characteristic cases of distance of the 
van-Hove peak to the Fermi level: $E_{vH}=30 meV$ (circles), $60 meV$
(squares) and $90meV$ (triangles).

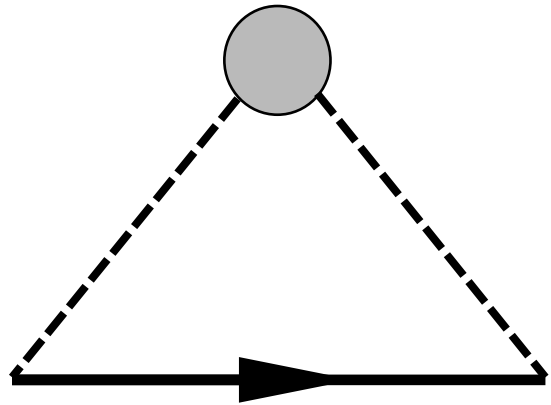
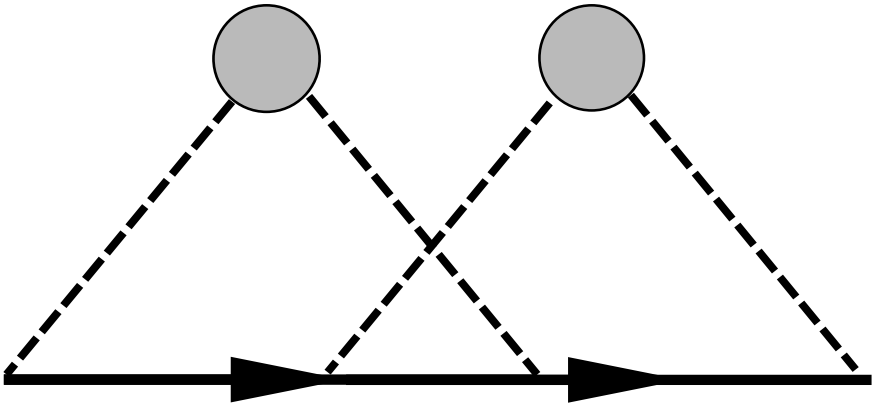
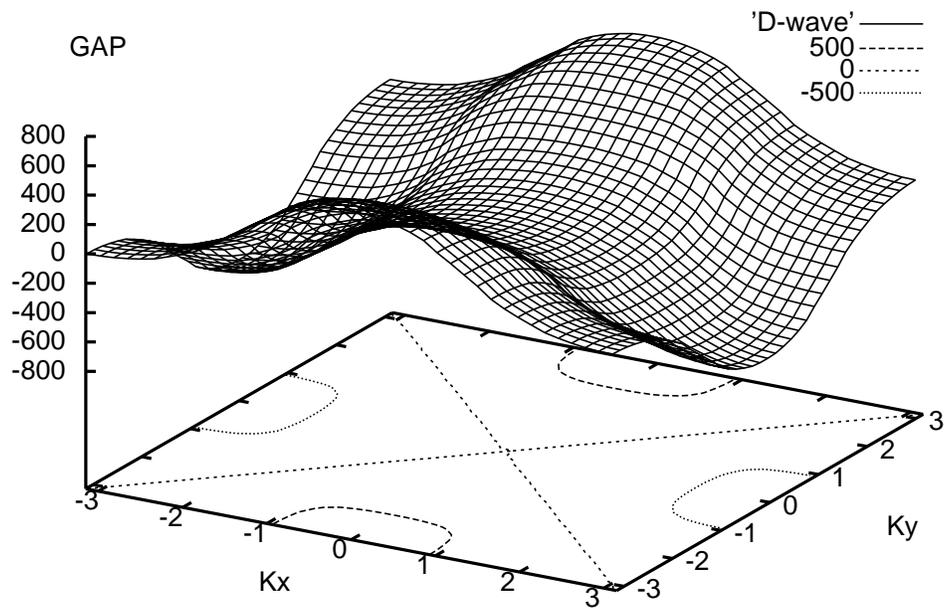
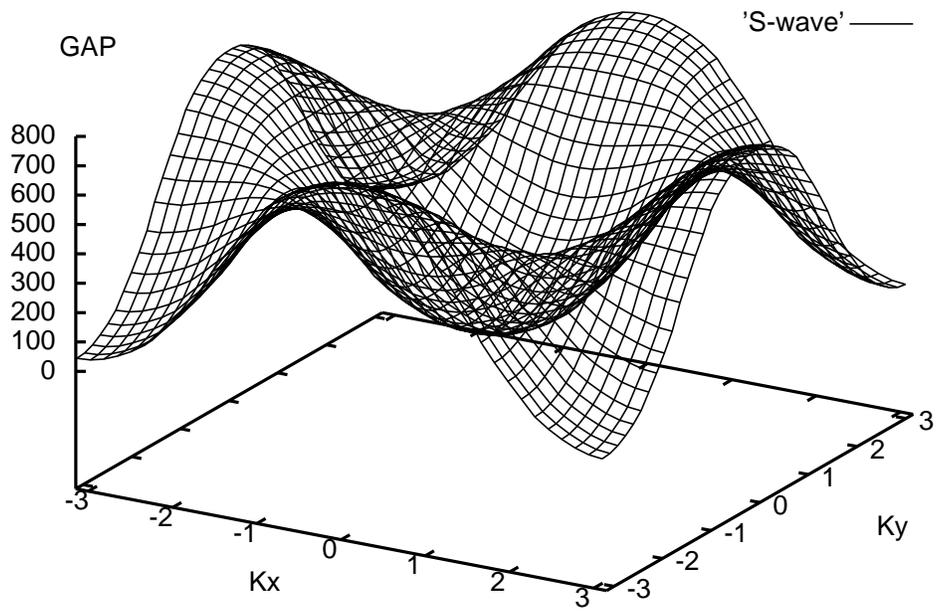
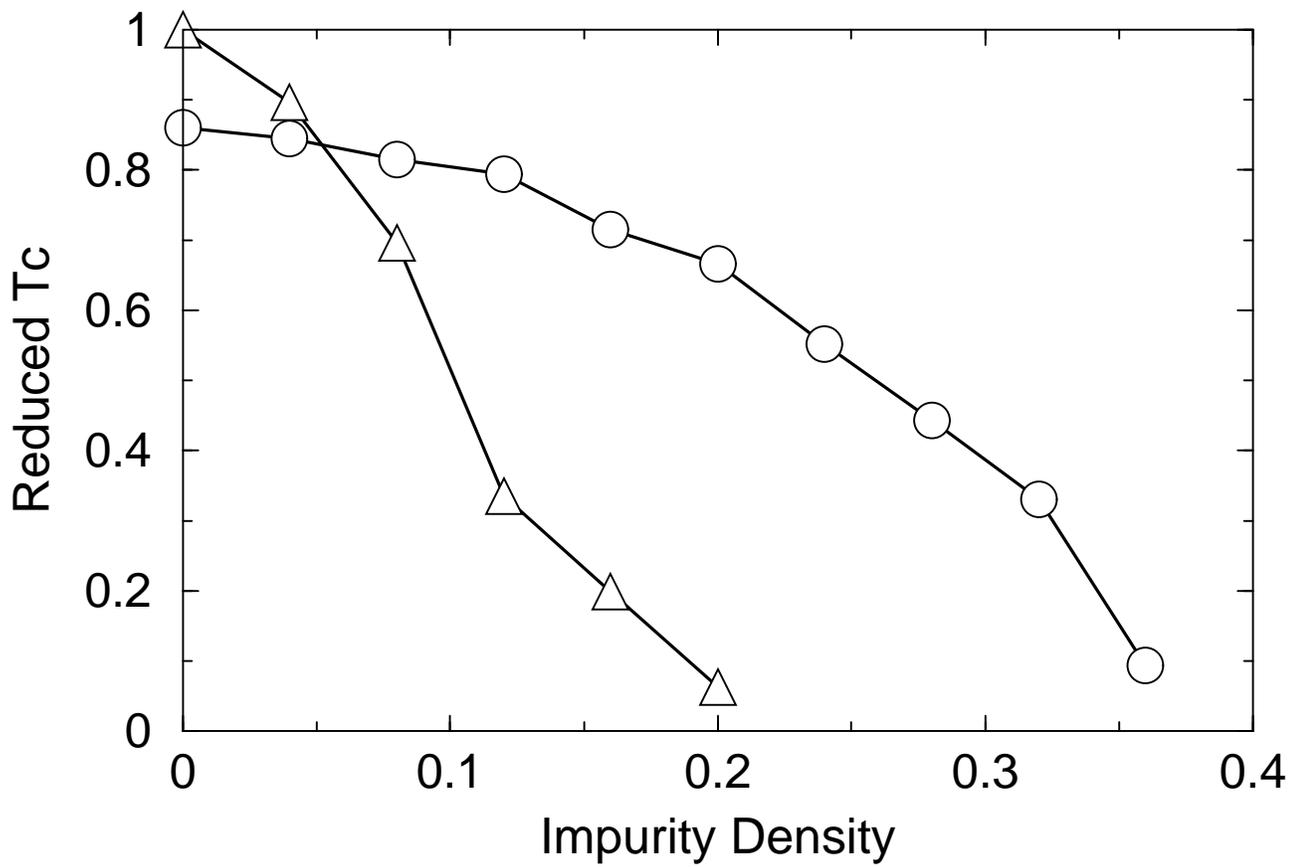
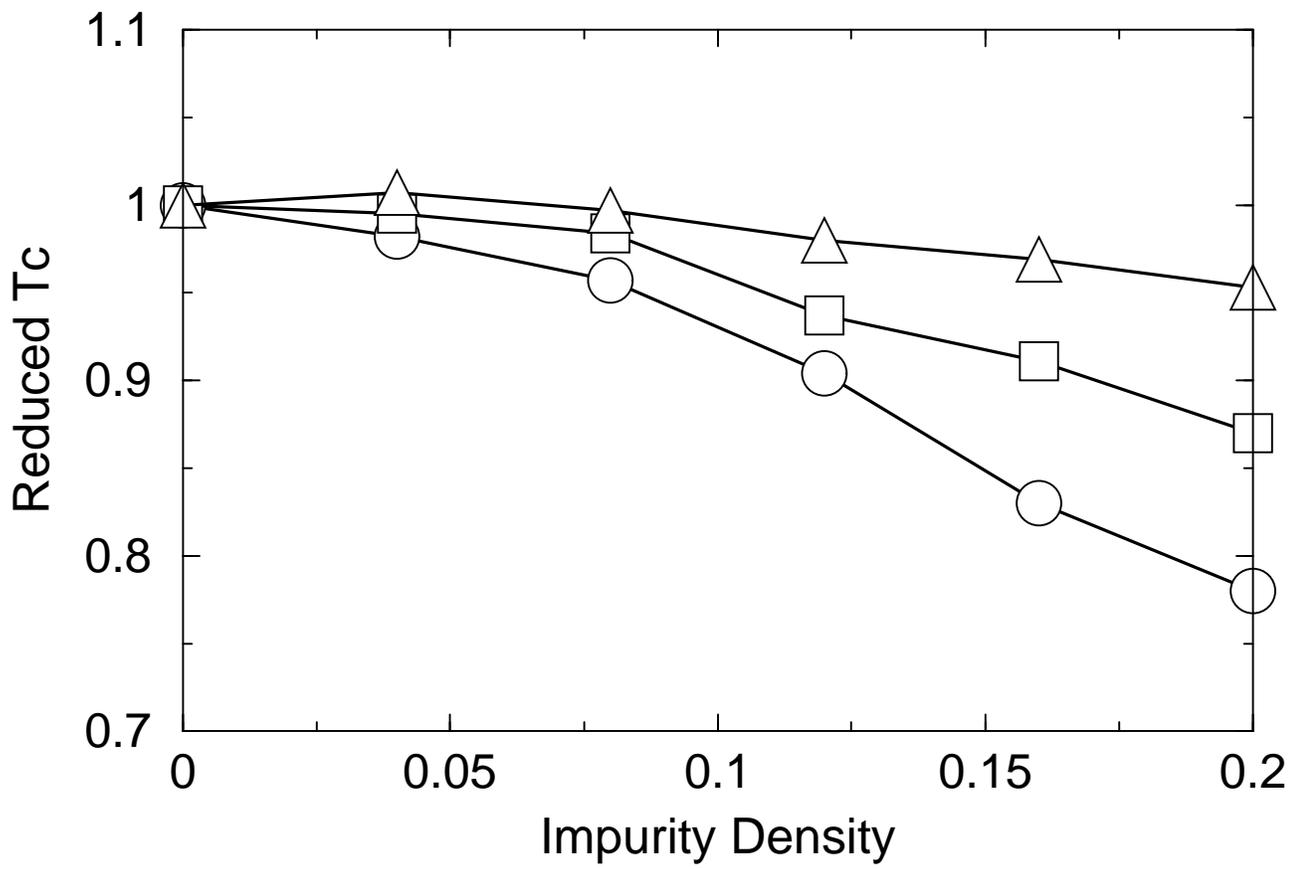
\end{document}